\documentclass[showpacs,prl,twocolumn,aps]{revtex4}
 \usepackage{bm}
 \usepackage{amsmath}
 \usepackage{amssymb}
 \usepackage{latexsym} 
 \usepackage{amsfonts} 
 \usepackage{epsfig}
 \usepackage{psfrag} 

 \newcommand{\kf}{{\boldsymbol k}}

 \newcommand{\bea}{\begin{eqnarray}}
 \newcommand{\ea}{\end{eqnarray}}
 \newcommand{\eea}{\end{eqnarray}}

\begin{document}

\title{Stability 
of quasi-two-dimensional 
Bose-Einstein condensates \\ with dominant dipole-dipole interactions}

\author{Uwe R.~Fischer}

\affiliation{Eberhard-Karls-Universit\"at T\"ubingen, 
Institut f\"ur Theoretische Physik\\  
Auf der Morgenstelle 14, D-72076 T\"ubingen, Germany}

\begin{abstract} 
We consider quasi-two-dimensional atomic/molecular Bose-Einstein condensates with both contact and dipole-dipole interactions.
It is shown that, as a consequence of the dimensional reduction, 
and within mean-field theory, the condensates do not develop unstable 
excitation spectra, 
even when the dipole-dipole interaction completely dominates 
the contact interaction. 
\end{abstract}

  
\maketitle

Dilute Bose-Einstein condensates with long-range interactions 
offer promising opportunities to explore  
the potentially strong
correlations induced by the interaction, which go beyond the
comparatively weak correlations 
induced by the local 
contact interaction pseudopotential conventionally 
sufficient to describe most atomic condensates. 
The recent realization of a concrete physical system 
in which atomic magnetic dipoles play a significant role \cite{Griesmaier}, 
is a first experimental step towards the exploration of dipolar condensate 
physics \cite{Goral,YiYouII,SantosZoller,ODell,Baranov}. 
To investigate {\em dipole-dominated} physics, greater potential than by 
magnetic dipoles is offered by heteronuclear molecules with an electric dipole
moment \cite{YiYou,Kotochigova}, due 
to the fact that the electric dipole
interaction strength is larger than the magnetic one by a 
factor of about $10^4$ for typical atomic/molecular 
dipole moments of 1 Bohr 
magneton or 1 Debye, respectively.  

An interesting property of three-dimensional (3D) 
condensates is the existence of a ``roton'' minimum, which stems from the 
fact that the Fourier transform of the dipole-dipole
interaction, occurring in the Bogoliubov mean-field spectrum,
assumes negative values in certain directions in momentum space
\cite{RotonDipolar,RotonsDipole,EPJD}.  
This property however also causes a problem 
of 3D dipolar condensates, as they become 
dynamically unstable for dipole-dipole interactions dominating the contact 
interaction, at densities corresponding to the Thomas-Fermi limit of 
large condensates. 
The roton minimum quickly deepens with increasing density and/or
dipole coupling to hit the zero of energy, and a further increase 
is impossible because the system develops imaginary excitation energies.

In the present study an analytical proof is given that, 
in contrast to 3D condensates,   
quasi-two-dimensional (quasi-2D) 
dipolar condensates are stable, 
even when the dipole coupling completely dominates the contact coupling. 
The experimental realization of purely dipolar condensates is thus 
possible only in the quasi-2D regime. The physical reason 
for the stabilization is that dimensional reduction entails 
that the mutual dipole-dipole interaction of the atoms/molecules, 
with dipoles oriented perpendicular to the
plane in the strongly confined direction, can effectively sample 
an exponentially smaller region in configuration space where the interaction
assumes negative values. 
It is, furthermore, shown that a 
strongly correlated regime of purely dipolar condensates, 
where large quantum depletion of the condensate occurs   
and the mean-field description in terms of a single condensate 
wave function breaks down, takes place in a  
crossover regime from quasi-2D to 3D.  
For an electric-dipole-dominated dilute gas, 
the densities corresponding to this crossover regime to 
3D turn out to be quite small, 
even within the strongly enlarged stability window of 
strongly anisotropic pancake-shaped condensates.

The mean-field description of the system we start with is based
on a nonlocal 3D Gross-Pitaevski\v\i\/ equation for the order
parameter $\Psi$ 
(we put $\hbar = m =1$, where $m$ is the mass of the atoms/molecules) 
\cite{Goral,YiYou}:  
\begin{eqnarray}
i \frac\partial{\partial t} \Psi ({\bm r}, t)
& = &  \left\{ -\frac{\nabla^2} 2 +\frac12 \omega_z^2 z^2 + 
g_{\rm 3D} |\Psi ({\bm r},t)|^2\right. \label{GP}\\
&  &\left. + \int d^3r' 
V_{dd}({\bm r}-{\bm r}') |\Psi ({\bm r}',t)|^2 \right\}\! \Psi ({\bm r}, t) .\nonumber
\end{eqnarray}
The trapping in the plane is assumed to be 
negligible as compared with the strong confinement, 
of harmonic trapping frequency $\omega_z$, in the $z$ direction. 
The term in the second line contains the dipole-dipole interaction of the 
atoms/molecules, with all dipoles oriented by an 
applied field along the strongly confining $z$ direction: 
\begin{equation} 
V_{dd} ({\bm r}) = \frac{3g_d}{4\pi} \frac{1-3z^2/|{\bm r}|^2}
{|{\bm r}|^3} . \label{Vdd}
\end{equation}  
The dipole coupling reads $g_d= \mu_0 d_m^2/3 $ for magnetic and 
$g_d=d_e^2/3\epsilon_0 $ for electric dipoles. In the units  
which we employ, the $g_d$ are having dimensions of length like 
the 3D contact interaction $g_{\rm 3D}=4\pi a_s$, for 
which the length scale is set by the $s$-wave scattering length $a_s$. 
We note that the value of the contact interaction coupling strength
 generally depends on the value of $g_d$ 
(in an effective-dimension dependent manner), 
because the long-range dipole-dipole interaction affects the 
short-range scattering processes \cite{YiYouII}. 
For the statements made in the present paper, 
the actual ratio $g_{\rm 3D}/g_d$ and the absolute 
value of the (unscreened) $g_d$ will be relevant. 

For a quasi-2D Bose-Einstein condensate, 
the motion of the atoms/molecules is 
by definition restricted to zero point oscillations
in a harmonic oscillator potential. 
We then take as a general ansatz for the density 
\begin{equation}
\rho({\bm r}) = |\Psi({\bm r})|^2 =\frac1{\sqrt{\pi d_z^2}} 
\exp\left[-\frac{z^2}{d_z^2}\right] n(x,y) ,
\label{quasi2D} 
\end{equation}
where the density in the plane, $n(x,y)$, is normalized to the 
total number of particles $N$. 
We treat $d_z$ as a parameter minimizing the 
Gross-Pitaevski\v\i\/ ground state energy \cite{YiYou,EPJD}.
Assuming homogeneous density in the plane, the equation determining 
$d_z$ reads $\omega_z^2= d_z^{-4} 
+ {(g_{\rm 3D} + 2g_d)} nN d_z^{-3}/{\sqrt{2\pi}}$. 
If the right-hand side is dominated by the first (kinetic energy) term, 
$d_z$  equals the harmonic oscillator length 
$1/\sqrt{\omega_z}$, the system is quasi-2D,  
and the above Gaussian gives 
the density profile exactly. 
Defining a parameter $\alpha \equiv \omega_z d_z^2$, we have 
$\alpha =1$ if the system is quasi-2D and $\alpha \gg 1$ 
deep into the 3D regime, where $\alpha$  becomes interaction dependent. 

We now calculate the total dipole-dipole energy given that the 
density profile in $z$-direction is prescribed by the above Gaussian.
In accordance with Eq.\,(\ref{GP}), 
the Gross-Pitaevski\v\i\/ energy functional 
of the dipole-dipole interaction generally reads   
\begin{eqnarray}
H_{dd} &= & \frac12 \int d^3r \int d^3r' \rho({\bm r}) V_{dd}({\bm r}-{\bm r}')
 \rho({\bm r}')  \nonumber\\
& = &  \frac12 \frac1{(2\pi)^3} \int d^3k \tilde \rho ({\bm k})
\tilde V_{dd} ({\bm k}) \tilde \rho (-{\bm k}),
\end{eqnarray} 
where the second line employs a convolution to Fourier space.  
The Fourier transform of the dipole-dipole interaction (\ref{Vdd})
takes the form 
$\tilde V_{dd} ({\bm k})=g_d [3k_z^2/(k_x^2+k_y^2+k_z^2)-1]$; 
using (\ref{quasi2D}), we obtain the Fourier transform of the density 
$\tilde \rho({\bm k})= \exp\left[-\frac14 k_z^2 d_z^2\right]\tilde n (k_x,k_y)$.  
Integrating over the $k_z$ direction, 
the effective dipole-dipole energy is then given by  
a two-dimensional integral in $(k_x,k_y)$ space 
\begin{eqnarray}
H_{dd} 
& = & \frac{g_d}2 \frac1{(2\pi)^2} \int d^2k \tilde n (k_x,k_y)
\tilde n (-k_x,-k_y) \nonumber\\ 
&  & \times \left\{ \frac{2}{\sqrt{2\pi} d_z} 
-\frac{3}2 \exp\left[\frac{k^2 d_z^2}2\right] k\, {\rm Erfc} \left(\frac{kd_z}{\sqrt 2}\right)\!
\right\} 
\label{Hdd}
\end{eqnarray} 
where the complementary error function 
Erfc $(z) = 1-{\rm erf}(z) = 1-(2/\sqrt{\pi}) \int_0^z \exp(-t^2) dt$
and $k=(k_x^2+k_y^2)^{1/2}$; $\tilde n(k_x,k_y)$ is the Fourier transform of 
the 2D density. 
Employing the same procedure of integrating out the Gaussian in the 
$z$ direction, for the contact interaction part in the 
Gross-Pitaevski\v\i\/ energy functional, yields the well-known result for 
the 2D effective coupling $g_{\rm 2D} = 2\sqrt{2\pi}
a_s/d_z$ (valid in the limit that the 3D $s$-wave scattering length 
$a_s\ll d_z$
\cite{Holzmann}).  
\begin{center}
\begin{figure}[t]
\psfrag{Vk}{\large $\tilde V^{\rm 2D}_{\rm tot}$} 
\psfrag{kz}{\large $kd_z$}
\centerline{\epsfig{file=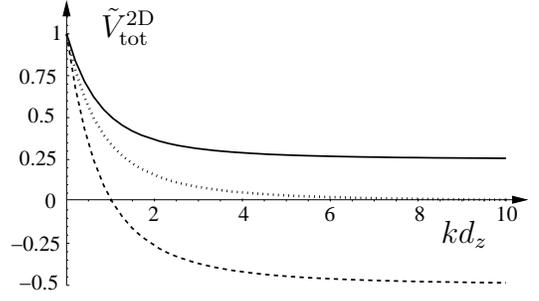,width=0.385\textwidth}}
\caption{\label{Fourier} 
Fourier transform of the total interaction potential 
in units of $A/\rho(0)\sqrt{\pi}d_z^3$, as a function of 
$\zeta =k d_z$. The values of $R=\frac12 \sqrt{\pi/2}$ (solid line), 
$R=\frac23 \sqrt{\pi/2} $ (dotted line)  
$R=\sqrt{\pi/2}$ (dipole interaction dominating; dashed line). For 
$R=\frac23 \sqrt{\pi/2}$ (i.e., when $g_{d}= g_{\rm 3D}$), 
the Fourier transform approaches zero at infinite
momentum.}
\end{figure}\vspace*{-2.5em}
\end{center}

From the relation (\ref{Hdd}) for the dipole-dipole contribution 
in the interaction energy, it follows that the Fourier transform of the 
total, contact plus dipole, interaction potential assumes the form 
\begin{equation} 
\tilde V^{\rm 2D}_{\rm tot} ({\zeta}) 
= \frac A{\rho(0)\sqrt{\pi} d_z^3}\left\{ 
1-\frac{3R}{2}\, \zeta\, w\left[\frac{\zeta}{\sqrt 2}\right]
\right\},  
\label{V2D}
\end{equation} 
where we made use of the $w$-function, related to Erfc$(z)$ by 
$w(z)= \exp[z^2]{\rm Erfc}(z)$ \cite{Abramowitz,note}.  
The dimensionless wavenumber ${\zeta} = k d_z $, and the two dimensionless parameters 
occurring in $\tilde V^{\rm 2D}_{\rm tot} ({\zeta})$, using the central
3D density $\rho(0)$, are defined to be  
\begin{equation}
A = \frac{\rho(0)\sqrt{\pi}d_z^2 g_d}R \,,\qquad 
R = \frac{\sqrt{\pi/2}}{1+g_{\rm 3D}/2g_d}.  \label{Adef} 
\end{equation} 
The value of the parameter $R$ ranges from $R=0$  
if $g_d/g_{\rm 3D} \rightarrow 0$, to $R=\sqrt{\pi/2}$ for 
$g_d/g_{\rm 3D}\rightarrow \infty$.
The second term in the curly brackets in Eq.\,(\ref{V2D})
rapidly decreases as a function of $\zeta$ and approaches a constant, 
which is due to the fact that $w(\zeta/\sqrt2) \rightarrow \sqrt{2/\pi} \, 
\zeta^{-1}$ for $\zeta\rightarrow \infty$, 
cf. the plot of $\tilde V^{\rm 2D}_{\rm tot} (\zeta)$ in  Fig.\,\ref{Fourier}. 
For small $\zeta$, the 
quasi-2D Fourier transform behaves like 
$\tilde V^{\rm 2D}_{\rm tot} ({\zeta}) 
= \frac A{\rho(0)\sqrt{\pi} d_z^3}
\left[1-\frac{3R}{2}{\zeta} +\frac{3R}{\sqrt{2\pi}}\zeta^2 +
{\cal O} (\zeta^3)\right]$. 
Observe that  $\tilde V^{\rm 2D}_{\rm tot} (\zeta)$ 
possesses a well-defined value at the origin $\zeta=k=0$, 
as opposed to the 3D Fourier transform of the
dipole-dipole interaction potential.

From the Fourier transform of the
interaction 
(\ref{V2D}), we conclude that the squared Bogoliubov spectrum 
\cite{Bogoliubov,Foldy}, 
for excitations confined to the plane, 
$\omega^2 = 
\rho(0)\sqrt{\pi} d_z \tilde V^{\rm 2D}_{\rm tot} (k) k^2 + k^4/4$ 
is, in units of 
$1/d_z^4$, given by 
\begin{equation} 
\epsilon^2(\zeta)=
A\zeta^2 \left( 1 - \frac{3R}2 
\zeta\,  w\left[\frac{\zeta}{\sqrt 2}\right]
\right)   
+ \frac{\zeta^4}4. \label{Bogoliubov} 
\end{equation}
The main observation of the present analysis is that, 
as opposed to the 3D case (or, in an exacerbated manner, 
the quasi-1D case \cite{EPJD}), the 
Bogoliubov spectrum (\ref{Bogoliubov}) does not necessarily become unstable 
if the dipole interaction coupling exceeds the contact interaction
coupling.
Assuming $g_{\rm 3D}/g_d \rightarrow 0$, i.e. $R\rightarrow\sqrt{\pi/2}$, 
the above squared spectrum can assume negative values, and hence the 
excitation energies become imaginary, when 
$A$ exceeds the critical value 
$A_c = 3.446 $, cf. Fig.\,\ref{Spectrum}, where we show a sequence of 
four spectra for different $A$ at constant $R=\sqrt{\pi/2}$.
  
The critical value of $A_c=3.446$ corresponds to a critical dipole coupling
given by  $({g_d)}_c = {2.436}/{\rho(0) d_z^2}$. 
Using numbers appropriate for atomic magnetic moments $\mu_m = N_m  \mu_B$, 
we find that $g_d = 5.4\times 10^{-6}\, \mu$m\,$M N_m^2$
($M$ is the mass of the atoms or molecules in units of the 
atomic mass unit $= 1.66\times 10^{-27}$\,kg). In the case
of electric dipoles, with moment $d_e= N_e$\,Debye,  
we find $g_d =  6.2\times 10^{-2} \,\mu$m\,$MN_e^2$.  
The critical coupling  $({g_d)}_c$ 
for dominant dipole-dipole interactions 
then translates into a critical central 
3D density 
$
\rho_c(0) =   4.5\times 10^{16} {\rm cm}^{-3} N_m^{-2} \alpha^{-1} 
\omega_z [2\pi \times
{\rm kHz}]
$  
and 
$
\rho_c(0) =   3.8\times 10^{12} {\rm cm}^{-3} N_e^{-2} \alpha^{-1}
\omega_z [2\pi \times {\rm kHz}] 
$
in the case of magnetic and electric dipoles, respectively. 
\vspace*{-1.25em}
\begin{center}
\begin{figure}[t]
\psfrag{E}{\large $\epsilon^2$}
\psfrag{kz}{\large $kd_z$}
\centerline{\epsfig{file=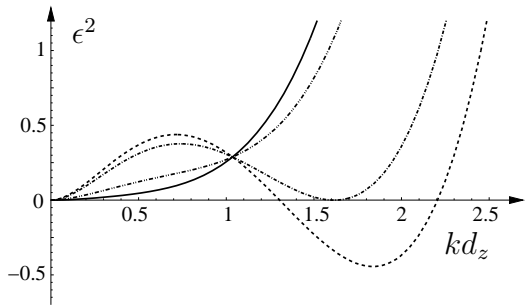,width=0.385\textwidth}}
\caption{\label{Spectrum} 
Squared Bogoliubov spectrum 
in units of $1/d_z^4$, Eq.\,(\ref{Bogoliubov}),  where 
$R=\sqrt{\pi/2}$ ($g_{\rm 3D}/g_d \rightarrow 0$) for all curves. 
Counting from bottom to top at small $kd_z$ the values of $A$ in 
Eq.\,(\ref{Adef}) are: 
$A=A_c/10$, $A=A_{\rm min}$, $A=A_c$, and $A= 1.2 A_c$. 
For $A>A_{\rm min}=1.249$, the spectrum displays a roton minimum, and becomes
unstable for $A>A_c=3.446$.} 
\end{figure}\vspace*{-1.5em}
\end{center} 

The quantity $\sqrt{\pi} \rho(0) g_d/R$, a measure of the 
energy per particle (the chemical potential), 
is equal to $A/d_z^2$. For the system to be 
quasi-2D, we therefore need $A\ll \omega_z d_z^2 =\alpha \equiv 1$. 
Furthermore, the Bogoliubov spectrum in Fig.\,\ref{Spectrum}  
does not possess any points where $d\epsilon^2/d\zeta^2$=0
for values  $A< A_{\rm min} = 1.249 $ (when $R=\sqrt{\pi/2}$). That is, 
a ``roton'' minimum cannot develop within the regime of quasi-2D.
Thus we conclude that a quasi-2D purely dipolar system of bosons is 
always stable, and that the instability of the 
condensate takes place in the crossover region to 3D. 
By contrast, deep inside the 3D regime, $g_{\rm 3D} \ll g_d$ 
implies collapse of the condensate for any reasonable value of the density,
and in the Thomas-Fermi limit 
the condensate will be unstable for any $g_d$ slightly exceeding $g_{\rm 3D}$
\cite{RotonDipolar}.
 
The critical value of $A_c(R)$, at which the gas becomes unstable, obtained 
by numerically finding the large momentum zeros of the Bogoliubov spectrum 
(\ref{Bogoliubov}), exponentially increases for smaller $R$ 
(increasing $g_{\rm 3D}/g_d$), and
diverges at $R=\frac23 \sqrt{\pi/2}$ (i.e., for $g_{\rm 3D}=g_d$),  
where the Fourier transform $\tilde V^{\rm 2D}_{\rm tot}$ becomes positive
everywhere. 
Comparing the critical values of $A$ thus obtained  
to those from a full 3D solution of the 
Bogoliubov-de Gennes equations \cite{RotonDipolar}, 
the present approach reproduces the critical $A$ sufficiently 
accurate in the dipole-dominated case,  
for which the instability takes 
place in the crossover regime from quasi-2D to 3D.  
Deep into the 3D regime, 
the critical value $A_c$ calculated from (\ref{Bogoliubov})
is strongly overestimated, mainly because
it is exponentially sensitive on the exact form of the spectrum.



At nonzero temperature $T$, 2D Bose-Einstein condensates
do not exist in the homogeneous case \cite{Hohenberg}, 
while trapping in the plane 
enables the existence of a condensate also at finite $T$ \cite{Fischer}. 
On the other hand, in the presently discussed $T=0$ case, 
it is a well-established fact 
that even without trapping, Bose-Einstein
condensation occurs in two spatial dimensions, 
see, e.g.,\,\cite{PitaLongRange}. 
We next turn to a discussion of the zero temperature value of the number
density of excitations above the condensate, the so-called quantum depletion.
To this end, we use a  
mode expansion for the  annihilation operators $\hat\chi_\kf$ 
of the original bosons in terms of the Bogoliubov 
quasiparticle operators $\hat a_{\bm k}, \hat a^\dagger_{\bm k}$ 
 \cite{Schutzhold}: 
\begin{equation}
\label{creation}
\hat\chi_{\boldsymbol k}
= \sqrt{\frac{{\boldsymbol k}^2}{2\epsilon_{\boldsymbol k}}}
\left[ \left(\frac12+\frac{\epsilon_{\boldsymbol k}}{{\boldsymbol k}^2}\right)
\hat a_{\boldsymbol k}
+\left(\frac12-\frac{\epsilon_{\boldsymbol k}}{{\boldsymbol k}^2}\right)
\hat a_{\boldsymbol k}^\dagger
\right].
\end{equation} 
The above form of the Bogoliubov transformation 
results, after inversion, in the usual phonon quasiparticle 
operators at low momenta, and gives 
$\hat \chi_{\bm k} = \hat a_{\bm k}$ at ${\bm k}\rightarrow \infty$, i.e.,  
the quasiparticles and the bare bosons become, as required, 
identical at large momenta. 

The quantum depletion density at zero temperature 
is calculated by evaluating the expectation of $\hat\chi^\dagger\hat\chi$ 
in the quasiparticle vacuum defined by 
$\hat a_{\bm k}|{\rm vac}\rangle=0$:
\bea
\label{dipole}
\langle 
\hat\chi^\dagger\hat\chi 
\rangle
&=& \frac1{2\pi d_z^2} \int_0^\infty d\zeta \, \frac{\zeta^3}{2\epsilon(\zeta)}
\left(\frac12 -\frac{\epsilon(\zeta)}{\zeta^2}\right)^2
.
\ea
The instability generally happens because {\em in-plane}  
excitations become of imaginary frequency \cite{RotonDipolar}; 
above the calculated critical value of $A=A_c$, 
the in-plane excitations have vanishing energy, and 
the quantum depletion 
diverges.    
Using the in-plane momentum integral above to calculate the depletion, with 
the effective dispersion relation (\ref{Bogoliubov}),  
assumes that the corrections due to the 
neglected out-of-plane Bogoliubov excitations,  
 characterized by transverse quantum numbers, 
are small.  This is justified because 
the dominant contribution to quantum depletion is, for a dilute Bose gas, 
from large momentum excitations with (approximately) 
vanishing energy.
Out-of-plane excitations have large energies at the relevant in-plane 
momenta of order $1/d_z$; they thus 
do not contribute significantly to the depletion.
In Fig.\,\ref{depl}, we show the result for the quantum 
depletion in the case of 
dipole-interaction  dominated condensates, for which we can expect the 
in-plane spectrum (\ref{Bogoliubov}) to be sufficiently 
accurate up to $A\simeq A_c$. 
At the critical value $A=A_c(\sqrt{\pi/2})\simeq 3.4$
(dotted vertical line in Fig.\,\ref{depl}), 
the condensate depletion diverges, and the mean-field 
condensate will yield to a new quantum phase.

\begin{center}
\begin{figure}[t]
\psfrag{Depl}{\normalsize $d_z^2\langle\hat\chi^\dagger\hat\chi\rangle$}
\psfrag{A}{\normalsize $A$}
\centerline{\epsfig{file=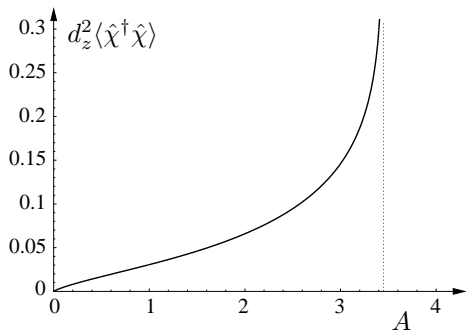,width=0.34\textwidth}}
\caption{\label{depl} 
Quantum depletion density from Eq.\,(\ref{dipole}),  
as a function of $A=\rho(0) \sqrt{\pi} d_z^2 g_d /R$,  
for $R=\sqrt{\pi/2}$ ($g_d/g_{\rm 3D} \rightarrow \infty$).}
\end{figure}
\vspace*{-2.5em}
\end{center}


In conclusion, we have shown that 
dipolar quasi-2D Bose-Einstein condensates are extremely  
stable systems as compared to their 3D counterparts. 
This offers the potential of  approaching,  starting
from the mean-field physics of condensates, a strongly correlated 
regime of dilute atomic/molecular gases with long-range interactions.
A conceivable experimental procedure is to 
start from a quasi-2D dipolar condensate, and, by 
decreasing $\omega_z$, to enter the crossover regime to 3D, 
where the quantum depletion of the condensate becomes large.   
If the dipole-dipole interaction dominates, i.e., in the limit $R\rightarrow
\sqrt{\pi/2}$, the quantity  
$A/\sqrt 2= \rho (0)d_z^2 g_d = 2.4 \,{\rho(0)}/{\rho_c(0)}$, 
where $\rho_c(0)$ is the critical 3D density for  
Bogoliubov excitations above the mean-field condensate to be stable, 
discussed in the above. Hence $\rho (0)d_z^2 g_d $ 
measures the diluteness of the system, that is the proximity to the 
critical region.
Numerical values are
\begin{eqnarray}
\rho (0)d_z^2 g_d & = &  
5.5 \times  10^{-3} \alpha N_m^2\, \frac{\rho(0) [10^{14} 
{\rm cm}^{-3}]}{\omega_z [2\pi\times {\rm kHz}]}\,, 
\nonumber\\
\rho (0)d_z^2 g_d  & = &  
0.63 \alpha N_e^2 \,\frac{\rho(0) [10^{12} 
{\rm cm}^{-3}]}{\omega_z [2\pi\times {\rm kHz}]}
\,,\label{diluteness} 
\end{eqnarray}  
in the magnetic and electric cases, respectively.
According to the first equation, 
magnetically dipolar gases will allow for the realization of a purely 
dipolar condensate, such that the depletion still remains small. 
On the other hand, it is evident from the second line of 
Eq.\,(\ref{diluteness}), that for electric dipoles  
the densities at which strong depletion of the condensate sets 
in are rather small, even for strongly increased axial trapping. 

Among further lines of research offered by the stability potential 
of quasi-2D dipolar condensates are the phenomena expected when they are
set in rotation. 
When both contact and dipole interaction are present, various of  
these phenomena have been explored 
in \cite{Cooper}; quantum Hall states of purely 
dipolar Fermi gases were studied in \cite{BaranovQH}. 
The interplay of quasi-two-dimensionality  
and rotation may generally 
yield interesting new physics
\cite{Sinha}.
It will, furthermore, be of interest to determine the change in the system's 
stability region when the dipoles are no longer locked in one direction,  
and thus to investigate the influence of spin waves on the stability of
dipolar 
quantum gases of bosons.

I thank C. Zimmermann, N. Schopohl,  L. Santos, 
C. Iniotakis, and N.\,R. Cooper for helpful discussions.


\begin{thebibliography}{499}
\bibitem{note} To find approximate analytical expressions for the spectrum, 
note that the $w$-function is to a good approximation bounded
from above and below by $(z+\sqrt{z^2+2})^{-1} < \frac{\sqrt{\pi}}2  
w(z)\le (z+\sqrt{z^2+4/\pi})^{-1} $ ($z\ge 0$) \cite{Abramowitz}.
\end{thebibliography}
\end{document}